\documentclass[twocolumn,amsmath,amssymb,prb,floatfix]{revtex4-1}

\usepackage{graphicx,color}
\usepackage{bmpsize}
\usepackage{dcolumn}
\usepackage{bm}
\usepackage{ulem}

\begin{document}

\title{Frustrated quantum magnetism with Bose gases in triangular optical lattices at negative absolute temperatures}

\author{Daisuke Yamamoto$^1$}
\email{d-yamamoto@phys.aoyama.ac.jp}
\author{Takeshi Fukuhara$^{2}$}
\author{Ippei Danshita$^{3}$}
\affiliation{$^1$Department of Physics and Mathematics, Aoyama Gakuin University, Sagamihara, Kanagawa 252-5258, Japan}
\affiliation{$^2$RIKEN Center for Emergent Matter Science (CEMS), Wako, Saitama 351-0198, Japan}
\affiliation{$^3$Department of Physics, Kindai University, Higashi-Osaka, Osaka 577-8502, Japan}
\date{\today}
\begin{abstract}
Quantum antiferromagnets with geometrical frustration exhibit rich many-body physics but are hard to simulate by means of classical computers. Although quantum-simulation studies for analyzing such systems are thus desirable, they are still limited to high temperature regions, where interesting quantum effects are smeared out. Here, we propose a feasible protocol to perform analog quantum simulation of frustrated antiferromagnetism with strong quantum fluctuations by using ultracold Bose gases in optical lattices at negative absolute temperatures. Specifically, we show from numerical simulations that the time evolution of a negative-temperature state subjected to a slow sweep of the hopping energy simulates quantum phase transitions of a frustrated Bose-Hubbard model with sign-inverted hoppings. Moreover, we quantitatively predict the phase boundary between the frustrated superfluid and Mott-insulator phases for triangular lattices with hopping anisotropy, which serves as a benchmark for quantum simulation.
\end{abstract}
\maketitle
Frustration is a key concept to understand various emergent phenomena in modern many-body physics~\cite{diep-04,moessner-06}. When different interactions among particles strongly compete with each other, e.g., for a geometric reason, the system is ``frustrated'' in determining the true ground state. The study on the interplay of the frustration and strong quantum fluctuations has been one of the core challenges of quantum many-body physics, presenting many open problems in connection with nontrivial magnetic states including quantum spin liquids~\cite{balents-10} and as a challenge for numerical techniques to handle highly entangled ground states~\cite{verstraete-08,kulagin-13}. {Quantum simulation with the use of ultracold atomic gases in optical lattices~\cite{lewenstein-07,bloch-12,gross-17}} has been discussed as a promising approach to make a critical breakthrough in this research area. {However, there still remain many challenges that have to be overcome in realizing and controlling frustrated quantum systems with cold atoms, whereas many theoretical proposals have been made~\cite{damski-05,ruostekoski-09,eckardt-10,huber-10,chen-10,yamamoto-19}}.

One straightforward idea for creating frustration is the use of two-component Fermi gases {with (pseudo)spins $\sigma=\uparrow,\downarrow$}~\cite{greif-13,hart-15,greif-15,parsons-16,boll-16,cheuk-16,mazurenko-17,hilker-17,brown-17} loaded into a nonbipartite (e.g., triangular~\cite{becker-10} and kagome~\cite{jo-12}) optical lattice. The second-order hopping process provides antiferromagnetic superexchange interactions between the (pseudo)spins {$\sigma$}~\cite{kuklov-03,duan-03}, which result in a frustrated situation because complete staggered spin configuration is not allowed by the lattice geometry. Although {long-range} magnetic correlation over a distance comparable to the system size has recently been observed in a square optical lattice~\cite{mazurenko-17}, a further technical breakthrough is required to realize far lower temperatures to study frustrated quantum magnetism. Another interesting idea is a fast shaking of optical lattice, which can effectively invert the sign of the hopping integral from the natural one~\cite{eckardt-05,lignier-07}. For ultracold Bose gases with sign-inverted hopping, the relative local phase of Bose-Einstein condensates (BECs) tends to be $\pi$ on neighboring sites, analogous to antiferromagnetic spin coupling, which induces geometric frustration in nonbipartite lattices. A frustrated classical XY model has been successfully simulated with this technique~\cite{struck-11}. However, it is {challenging} to reach a quantum regime of low temperature and density because the lattice shaking can be a source of heating.

{Recently, it has become realistic to create a well-controlled system at negative absolute temperatures~\cite{ramsey-56} in laboratory.} A state in thermal equilibrium is usually described by a statistical ensemble in which the lower-energy states are more occupied than higher-energy ones, obeying the probability proportional to the Boltzmann factor with temperature $T\geq 0$. However, if the system has an upper energy bound, the opposite distribution with the largest occupation of the highest energy could also manage to be prepared. Such a state is characterized by a negative absolute temperature $T\leq 0$. In the pioneer work of Ref.~\cite{braun-13}, Braun $et$ $al$. have created a thermodynamically-stable negative-temperature state of Bose gases in a square optical lattice by achieving the maximum interaction and potential energies in the regime of negligible kinetic energy. There the absolute temperature remains so low that the quantum phase transition from the Mott insulator (MI) to the superfluid (SF) has been observed.

{Here we propose and examine a realistic route to create frustrated Bose gases in a quantum regime using the combination of the phase-imprinting techniques~\cite{dobrek-99,burger-99,taie-15} and the statistics of {negative} absolute temperatures. Our proposal} is based on the fact that a negative-temperature state of a system $\hat{\mathcal{H}}$ at $T<0$ realizes the corresponding equilibrium state of {\it the sign-inverted Hamiltonian} $-\hat{\mathcal{H}}$ at $|T|>0$. Using this, one can achieve the same effect as inverting the sign of hopping, instead, by inverting the other factors, namely interatomic interaction and trap potential, and then preparing a negative-temperature state. To this end, we propose a phase-imprinting scheme combined with sudden sign inversion of the interaction and potential, which causes much less heating compared to the lattice shaking{. Supposing} Bose gases in a triangular lattice, we simulate the dynamics along the protocol within the time-dependent Gutzwiller approach (TDGA) to demonstrate that quantum phases of the frustrated Bose-Hubbard model, including chiral superfluid (CSF), could indeed be realized. {Moreover, considering more general hoppings with spatial anisotropy, in which the hopping amplitude in one direction can be different from those in the other two directions}, we give more quantitative analysis on the quantum phase transition between frustrated CSF and MI by means of the cluster mean-field plus scaling (CMF+S) method~\cite{yamamoto-12-2,yamamoto-14,yamamoto-17} with two-dimensional (2D) density matrix renormalization group (DMRG) solver~\cite{yamamoto-19}. This enables us to discuss the interplay of frustration and quantum fluctuations, which is a critical factor for producing various exotic frustrated states. The theoretical predictions provide a solid guidepost for future experiments to confirm that the interplay effects are properly captured in the quantum simulator. We set $\hbar=1$ except in the figures.
\begin{figure}[t]
\includegraphics[scale=0.4]{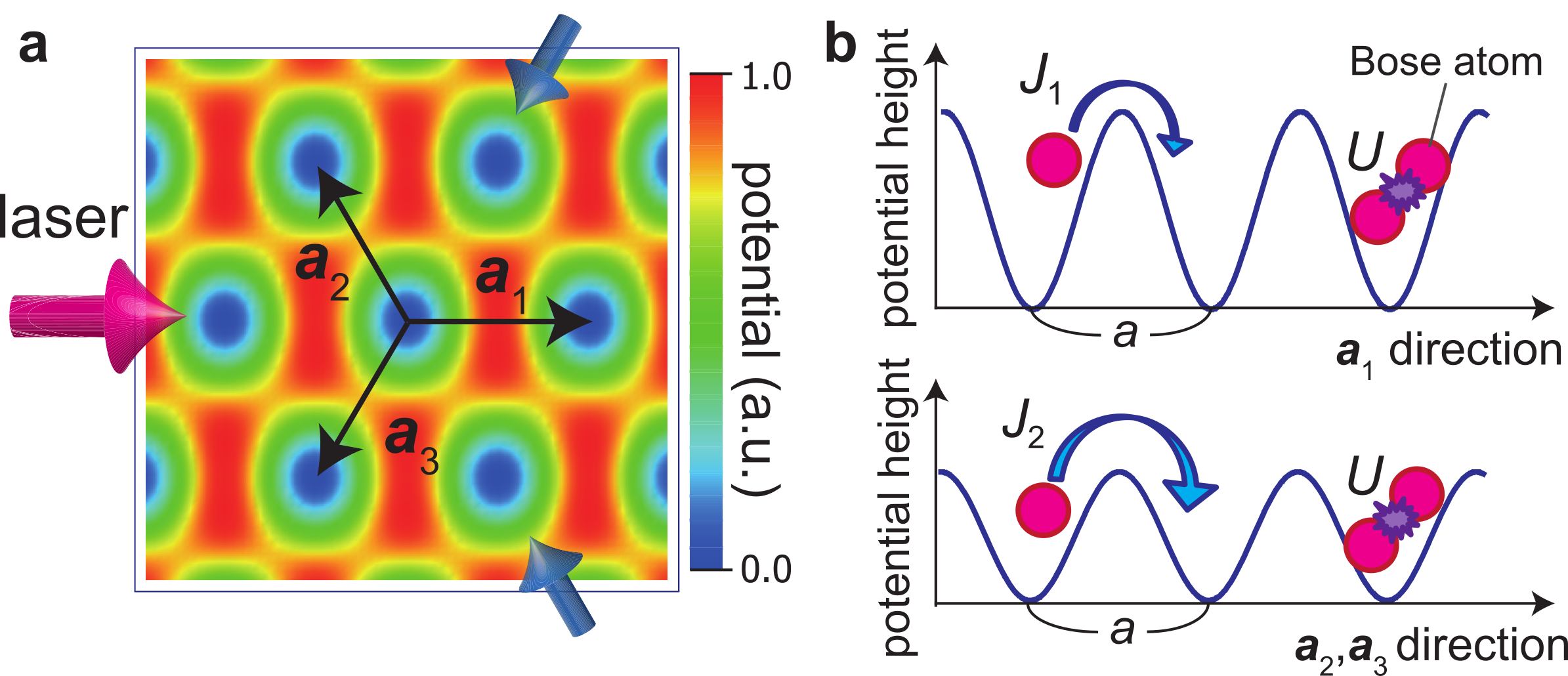}
\caption{\label{fig1}
{\bf Triangular optical lattice with anisotropic hoppings.} {(a) Typical potential landscape of isosceles triangular optical lattice created by three laser beams in the case where one of the three lasers (in the ${\bf a}_1=(a,0)$ direction with $a$ being the lattice constant) has a larger intensity than the other two (in the ${\bf a}_2=(-a/2,\sqrt{3}a/2)$ and ${\bf a}_3=(-a/2,-\sqrt{3}a/2)$ directions). (b) Schematic figures of the modeling with Bose-Hubbard parameters: the hoppings $J_1$ in the ${\bf a}_1$ direction and $J_2$ in the ${\bf a}_2$ and ${\bf a}_3$ directions, and the onsite interparticle interaction $U$. The size of the arrows schematically shows the difference of the hopping amplitudes depending on the potential height between the neighboring lattice sites in each direction. } }
\end{figure}
\\ \\    
{\bf \large Results}\\
{\bf The Bose-Hubbard model on triangular lattice.} 
A system of Bose gases in a deep optical lattice is described by the Bose-Hubbard model: 
\begin{eqnarray}
\!\!\!\!\!\!\hat{\mathcal{H}}=-\sum_{i,j}J_{ij}\hat{b}_{i}^\dagger\hat{b}_{j}+\frac{U}{2}\sum_{i}\hat{n}_{i}(\hat{n}_{i}-1)+\frac{V}{a^2} \sum_i |{\bf r}_i|^2 \hat{n}_{i}\label{hamiltonian}
\end{eqnarray}
with hopping integral $J_{ij}$ {(for $i \neq j$), chemical potential $J_{ii}\equiv \mu$}, onsite interaction $U$, harmonic trap potential $V |{\bf r}_i|^2/a^2$, and lattice constant $a$. {Here we consider a triangular optical lattice with spatially anisotropic hopping of ``isosceles type'', which is parameterized by $J_{ij}=J_1$ for nearest neighbor (NN) sites $(i,j)$ in the ${\bf a}_1=(a,0)$ direction and $J_{ij}=J_2$ in the ${\bf a}_2=(-a/2,\sqrt{3}a/2)$ and ${\bf a}_3=(-a/2,-\sqrt{3}a/2)$ directions [see Fig.~\ref{fig1}]. The spatial anisotropy can be created by tuning the intensity of one of the three lasers to be different from the others (see Methods).} 
The two extreme limits, $J_1/J_2=0$ and $J_1/J_2\gg 1$, are reduced to square lattice and 1D chain, respectively. 
\\\\
{\bf Gutzwiller analysis.}  We first discuss the ground state for $V=0$ within the site-decoupling Gutzwiller approach (GA)~\cite{GA1,GA2,buonsante09} to get a basic insight into the problem. {The ground state in the weak-coupling regime ($U\ll |J_{ij}|$) is well described within the GA under the assumption of the BEC order {$\langle b_i\rangle\equiv \psi_i= \bar{\psi} e^{i{\bf q}\cdot {\bf r}_i+\varphi}$} with momentum ${\bf  q}$ and global phase $\varphi$. The} kinetic energy of Eq.~(\ref{hamiltonian}) is given by {$-\sum_{i,j\neq i}J_{ij}\langle\hat{b}_{i}^\dagger\hat{b}_{j}\rangle\approx \varepsilon_{{\bf q}}\bar{\psi}^2M$}. Here, $\varepsilon_{{\bf q}}\equiv -2(J_1\cos {\bf q}\cdot {\bf a}_1+J_2\cos {\bf q}\cdot {\bf a}_2+J_2\cos {\bf q}\cdot {\bf a}_3)$ and $M$ is the number of lattice sites. For natural-sign hoppings $J_1,J_2>0$, the minimum kinetic energy is obtained at ${\bf q}={\bf 0}$, leading to a uniform SF state. The maximum of the kinetic energy is achieved at ${\bf q}=\pm {\bf Q}$ with
\begin{eqnarray}
{\bf Q}=\left\{ \begin{array}{ccc} (2\pi/a,0){\equiv{\bf Q}_{\rm M}}&{\rm for}&0\leq J_1/J_2 \leq 0.5,\\ (2\arccos[\frac{-J_2}{2J_1}]/a,0)&{\rm for}&J_1/J_2 > 0.5.\end{array} \right. \label{BECmomentum}
\end{eqnarray}
Therefore, if sign-inverted hopping ($J_1,J_2<0$) is prepared, a frustrated CSF state with finite BEC momentum ${\bf q}=\pm {\bf Q}$ is realized. The choice of ${\bf q}={\bf Q}$ or $-{\bf Q}$ represents the degeneracy with respect to the chirality of vortex in unit triangles. 
Hereafter we suppose ${\bf q}={\bf Q}$ to be spontaneously selected.  
{In the equilateral case ($J_1=J_2\equiv J$), the momentum is ${\bf Q}=(4\pi/3a,0)\equiv{\bf Q}_K $. Therefore, the CSF state} forms a ``three-color'' arrangement of the local phase (${\rm Arg}[\psi_i]=0$, $2\pi/3$, and $4\pi/3$ within a global phase shift) as shown in Fig.~\ref{fig2}a. This can be understood as a compromise solution for the frustration in the bond energy minimization. For generic $J_1/J_2 > 0.5$, the phase factor changes spatially with incommensurate pitch vector ${\bf Q}$. In the parameter range $0\leq J_1/J_2\leq 0.5$ and the 1D limit $J_1/J_2\gg 1$, a ``two-color'' ($0$ and $\pi$) pattern is formed with no chiral degeneracy. The $J_1/J_2$ dependence of ${\bf Q}$ is analogous to the pitch vector of spin spiral states in spatially-anisotropic triangular antiferromagnets~\cite{spinwave1,spinwave2,coldea-01,zvyagin-19}.
\begin{figure}[b]
\includegraphics[scale=0.4]{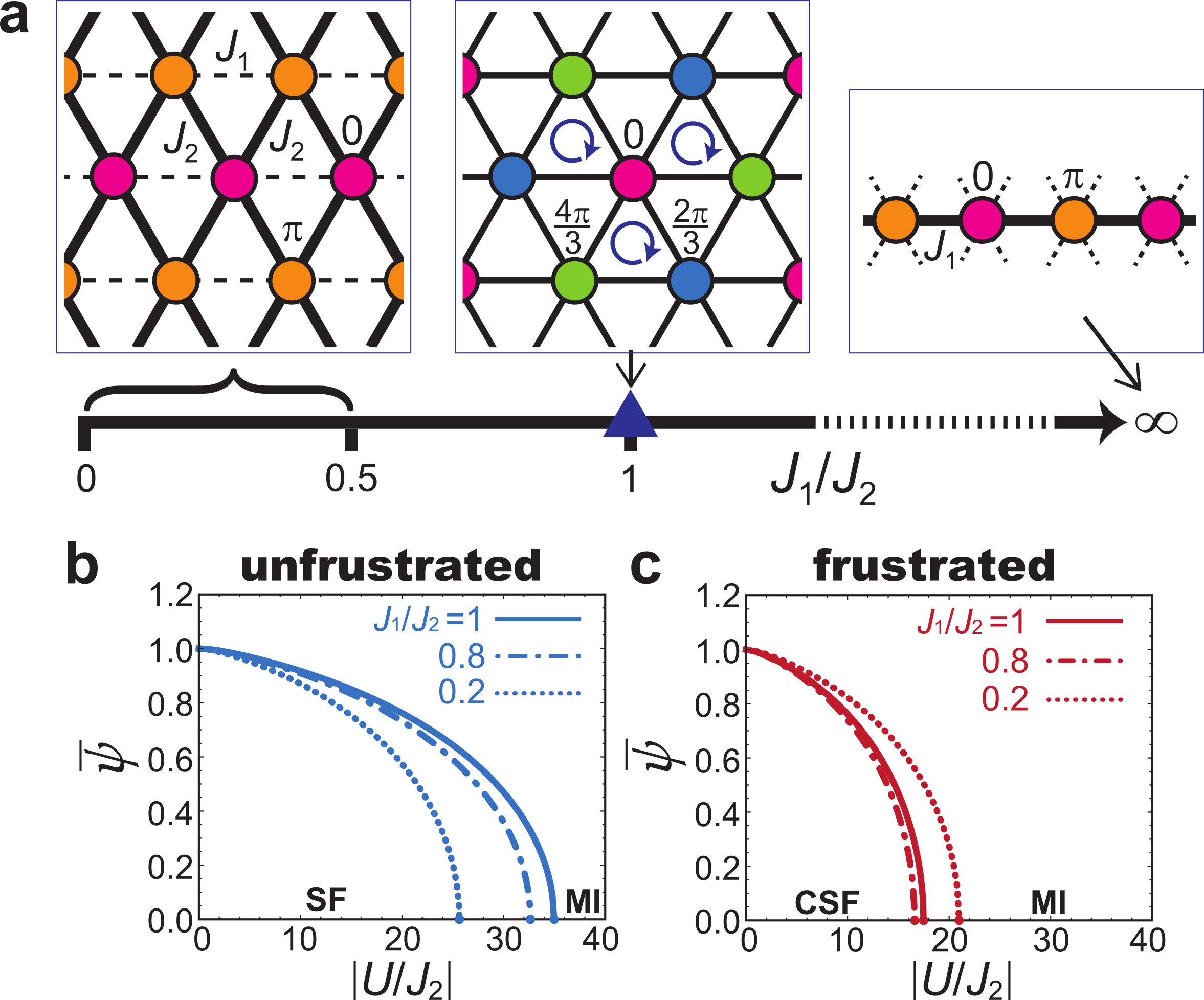}
\caption{\label{fig2}
{{\bf Predictions from the Gutzwiller approach.} (a) The $J_1/J_2$ dependence of the local phase pattern in the frustrated case $J_1,J_2<0$ is illustrated. The colors indicate the sublattice structure in the local phase and the arrows depict the chirality. (b,c) The order parameter {$\bar{\psi}$} as a function of $|U/J_2|$ (being the ratio of the interaction and the hopping in the ${\bf a}_2$ and ${\bf a}_3$ directions) at the filling factor $\rho = 1$ for (b) $J_1,J_2>0$ and (c) $J_1,J_2<0$. The value of $\bar{\psi}$ becomes zero at the transition from the superfluid (SF) or chiral superfluid (CSF) to Mott insulator (MI) phase. } }
\end{figure}
\begin{figure*}[t]
\includegraphics[scale=0.50]{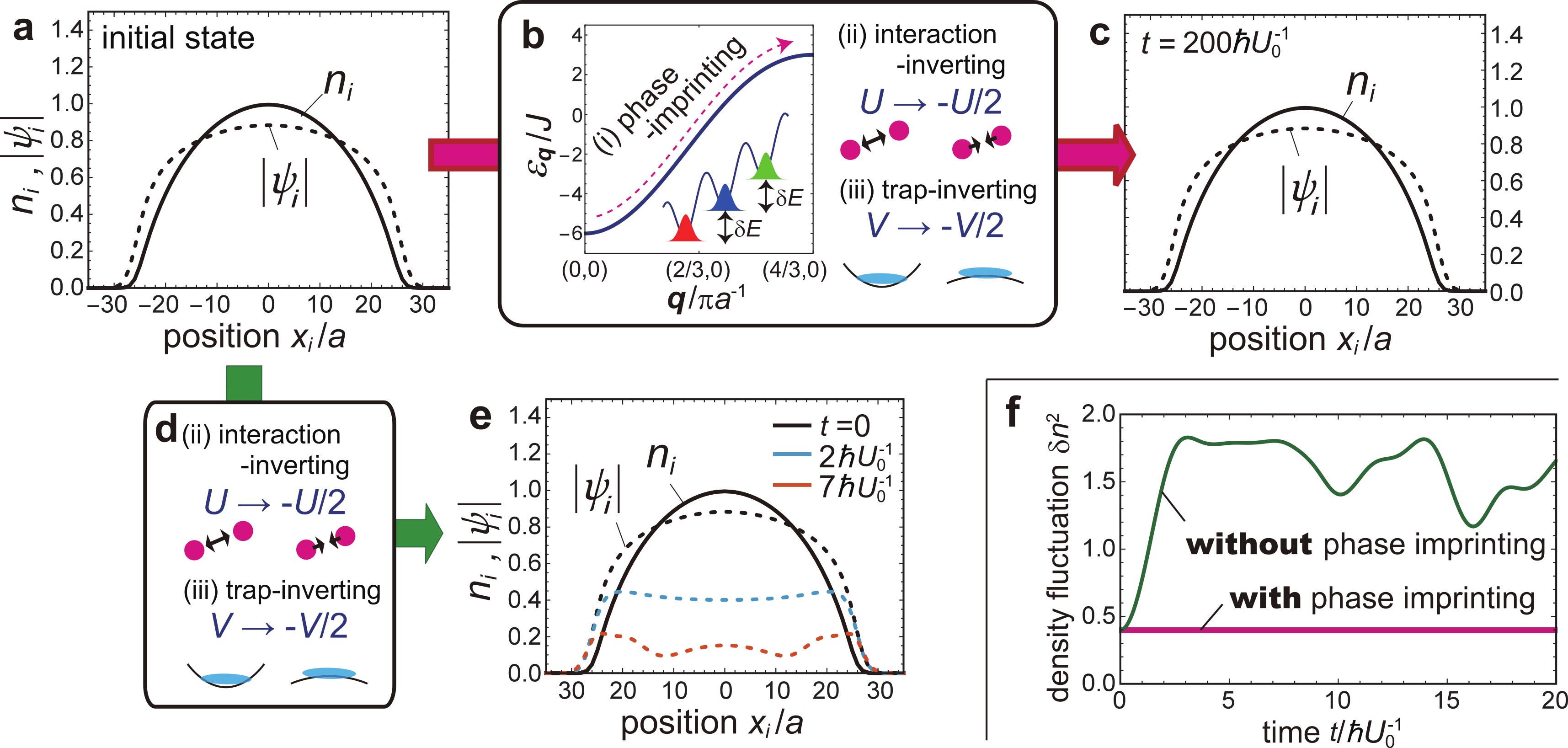}
\caption{\label{fig3}
{\bf Dynamical stability of the negative-temperature {chiral superfluid} (CSF) state within the {time-dependent Gutzwiller approach}.} (a) Profiles of density {$n_i$} (solid lines) and order parameter {$|\psi_i|$} (dashed lines) in the harmonic trap along the cut of $y_i=a/\sqrt{3}$ {(with $a$ being the lattice constant)} in the initial state. (b) The three operations to make frustrated negative-temperature states (for {spatially isotropic hoppings} $J_1=J_2=J$). {The single-particle spectrum $\varepsilon_{{\bf q}}$ in units of $J$ is plotted as a function of the wave vector ${\bf q}$ in units of $\pi a^{-1}$. The illustrations depict a tilting of the optical lattice by energy difference $\delta E$ for the phase-imprinting, the inverting from replusive ($U>0$) to attractive ($-U/2<0$) interaction, and the inverting from confinement ($V>0$) to anti-confinement ($-V/2<0$) trap potential, respectively}. (c) {Negative-temperature CSF state after time $200$ {in units of the inverse of the initial interaction strength $U_0^{-1}$}, which shows its stability for a sufficiently long time}. The case without the phase imprinting operation (d) is shown in (e) for a comparison. (f) The evolution of the density fluctuation $\delta n^2$ of the initial state with and without the phase imprinting.
}
\end{figure*}

For large, repulsive interaction $U$, lattice bosons undergo a quantum phase transition to the MI state when the filling factor {$\rho\equiv M^{-1}\sum_i\langle\hat{n}_i\rangle$} is an integer~\cite{greiner-02}. Performing the GA decoupling $\hat{b}_i^\dagger\hat{b}_j\approx \psi_j\hat{b}_i^\dagger+\psi_i^\ast \hat{b}_j-\psi_i^\ast\psi_j$ in Eq.~(\ref{hamiltonian}), we calculate the order parameter $\bar{\psi}$ for $\rho=1$ in the unfrustrated ($J_1,J_2>0$: ${\bf q}={\bf 0}$) and frustrated ($J_1,J_2<0$: ${\bf q}={\bf Q}$) ground states {as a function of $U/J_2$} (see Methods). {Figures~\ref{fig2}b and~\ref{fig2}c show the GA results for the SF-to-MI and CSF-to-MI quantum phase transitions, respectively}. At $J_1=J_2= J$, the critical point at which {$\bar{\psi}$} vanishes is given as $U^{({\rm GA})}_{\rm c}/|J|=17.5$ for the frustrated case, which is a half of $U^{({\rm GA})}_{\rm c}/J=35.0$ for the unfrustrated case. The strong reduction is attributed by the fact that the CSF state is less stable due to the frustrated local-phase arrangement {in which the NN bonds are not fully satisfied in minimizing the local kinetic energy}. For general {values of the anisotropy} $J_1/J_2$ and $\rho$, the critical point is given by $U^{({\rm GA})}_{\rm c}= -\varepsilon_{{\bf q}}(\sqrt{\rho}+\sqrt{\rho+1})^2$ with ${\bf q}={\bf 0}$ (${\bf q}={\bf Q}$) for the unfrustrated (frustrated) case. The ratio $|\varepsilon_{\bf{Q}}/\varepsilon_{\bf{0}}|$ equals to 1 only at $J_1/J_2=0$ or {$J_1/J_2\rightarrow \infty$, indicating that} the reduction effect due to frustration exists even in the ``two-color'' region of $0<J_1/J_2\leq 0.5$.

{It is noted that the curves of $\bar{\psi}$ for the unfrustrated system in Fig.~\ref{fig2}b exactly overlap those for the frustrated system {in Fig.~\ref{fig2}c}, respectively for each $J_1/J_2$, by changing the scale of the interaction $U$ by the factor $|\varepsilon_{\bf{Q}}/\varepsilon_{\bf{0}}|$ (specifically, the factor 1/2 when $J_1=J_2$). This is the case for general values of $J_1/J_2$ and $\rho$. Indeed, the ground-state properties, including the density and order parameter, for the unfrustrated system with certain $J_{ij},U>0$ and those for the frustrated system with negative hopping $-J_{ij}$ and repulsion $|\varepsilon_{\bf{Q}}/\varepsilon_{\bf{0}}|U$ are identical within the GA, except for the spatial phase distribution ${\rm Arg}[\psi_i]={\bf Q}\cdot {\bf r}_i$. This is because the scale of the kinetic energy for the frustrated system is reduced by the factor $|\varepsilon_{\bf{Q}}/\varepsilon_{\bf{0}}|$ due to the frustrated phase configuration (see Methods).}
\\ \\{\bf Negative absolute temperature.} 
To experimentally create the frustrated quantum states, it is required to prepare sign-inverted hoppings $J_{ij}<0$ with avoiding serious heating of the system. Below, we explain the details of the protocol through use of negative-temperature statistics. 
Let us suppose an initial state in which $N$ particles are distributed in a triangular lattice and a harmonic potential, which realizes the SF ground state of the standard Bose-Hubbard model (\ref{hamiltonian}) with $J_{ij}>0$ and $U,V>0$. See a typical example in Fig.~\ref{fig3}a obtained within the GA~\cite{buonsante09} for $N=1400$, $J_1=J_2=J=0.08U_0$, $U=U_0$, and $V=0.001U_0$ (with $U_0>0$ being the energy unit). 
{The phase ${\rm Arg}[\psi_i]$ is uniform in the unfrustrated SF state. }

{First, one needs to introduce the spatial phase distribution ${\rm Arg}[\psi_i]={\bf Q}\cdot {\bf r}_i$ to create the frustrated CSF state.}
To this end, here we suggest the use of the phase-imprinting techniques~\cite{dobrek-99,burger-99,taie-15}. When a single-particle energy difference $\delta E$ is introduced between two sites, the relative phase on the two sites starts to develop with $\exp [i \delta E t]$ in time $t$. In a region deep inside the SF phase, the kinetic energy ($\propto \varepsilon_{{\bf q}}$) reaches the maximum when the BEC momentum takes ${{\bf q}}={{\bf Q}}$ given in Eq.~(\ref{BECmomentum}). One can transfer ${\bf q}$ from ${\bf 0}$ to ${\bf Q}$ by introducing a temporary linear gradient potential $\hat{V}^{\rm ext}= \delta E \sum_i(x_i/a) \hat{n}_i$ for appropriate time $\delta t=Q_x a/\delta E $ [see Fig.~\ref{fig3}b]. Such a temporary potential can be created, e.g., by a magnetic field gradient or by an extra 1D optical lattice satisfying ${\bf Q}={\bf Q}_{\rm M}$ or ${\bf Q}={\bf Q}_{\rm K}$ (see Methods). 
{One has to perform the phase-imprinting operation in a much shorter time than the time scale of the hopping ($\delta t\ll |J_{ij}|^{-1}$) in order to affect only the local phases. Besides, the local energy difference $\delta E$ has to be set to a large enough value compared to $|U|$ and $|V|$ so that one can safely avoid the influence of the inhomogeneity of the density profile (see Methods). Since such well-controlled phase imprinting has been successfully made in previous experiments~\cite{burger-99,taie-15}, we will assume in the theoretical simulations presented below that a perfect phase imprinting can be achieved. The created CSF state with the ``forced'' phase distribution $e^{i{\bf Q}\cdot {\bf r}_i}$ should of course be dynamically unstable, since it has the maximum kinetic energy.}

{By changing $U$ and $V$ to be attractive ($U<0$) and anti-trapping ($V<0$), we make the interaction and potential energies also reach their maximum, in order to realize a stable negative-temperature ground state (at $T\approx -0$) for $J_{ij}>0$ and $U,V<0$. The created negative-temperature state should be equivalent to the ground state (at $|T|\approx +0$) of the frustrated system with $J_{ij}<0$ and $U,V>0$ since the physics of the two systems obey the same factor $\exp [-\hat{\mathcal{H}}/k_{\rm B}T]$.} 
In Ref.~\cite{braun-13} for square (unfrustrated) lattice, this has been performed simply with $U\rightarrow -U$ and $V\rightarrow -V$ by using the Feshbach resonance~\cite{chin-10} and blue-detuned lasers. {In the present case with frustration, one has to pay special attention to the change of the kinetic energy scale after the sign inversion of $U$ and $V$. Specifically, the initial state with the density ($n_i$) and order-parameter ($|\psi_i|$) distributions shown in Fig.~\ref{fig3}a after the phase imprinting ${\rm Arg}[\psi_i]={\bf Q}\cdot {\bf r}_i$ is expected to correspond to the ground state of the frustrated Hamiltonian with the interactions rescaled by the factor $|\varepsilon_{\bf{Q}}/\varepsilon_{\bf{0}}|$, as explained above in the homogeneous case. Therefore,} the values of $U$ and $V$ have to be changed as $U\rightarrow -|\varepsilon_{\bf{Q}}/\varepsilon_{\bf{0}}|U$ and $V\rightarrow -|\varepsilon_{\bf{Q}}/\varepsilon_{\bf{0}}|V$ (e.g., $U\rightarrow -U/2$ and $V\rightarrow -V/2$ for $J_1=J_2$) in order to adjust the energy scale in consideration of the reduction of the kinetic energy due to the frustration. 
\\ \\ {\bf TDGA simulation.}--- 
{In the framework of the TDGA~\cite{damski-03,zakrzewski-05}, we simulate the time evolution of the initial state in Fig.~\ref{fig3}a after suddenly performing the three operations shown in Fig.~\ref{fig3}b. In the simulation, we implement the phase imprinting on the initial state by the operation $\sum_n f_i^{(n)}|n\rangle \rightarrow \sum_n e^{in{\bf Q}_{\rm K}\cdot{\bf r}_i } f_i^{(n)}|n\rangle$ on the local wave function at every site $i$. {Here, $f_i^{(n)}$ is the coefficient of the Fock basis $|n\rangle$.} In addition, we perform the sudden changes of $U=U_0\rightarrow -U_0/2$ and $V=0.001U_0\rightarrow -0.0005U_0${. After those three operations are performed at $t=0$, we calculate the time evolution of the state fixing the value of $|U/J|$ for $0<t \leq 200 U_0^{-1} $ to see the stability of the created negative-temperature CSF state. }

As shown in Fig.~\ref{fig3}c, the created negative-temperature state is predicted to be indeed dynamically stable for a long time $0<t \leq 200 U_0^{-1} $. {As a reference for the comparison, we also simulate the case with the same settings but without the phase-imprinting operation [Fig.~\ref{fig3}d]. In this case, as shown in Fig.~\ref{fig3}e}, the state collapses immediately within $t \lesssim 2 U_0^{-1} $ due to the dynamical instability; the order parameter rapidly decreases, although the density profile is kept. The difference between the two cases (with and without the phase imprinting) can be clearly seen in the time evolution of the density fluctuation $\delta{n}^2\equiv \overline{\langle\hat{n}_i^2 \rangle-\langle\hat{n}_i \rangle^2}$ averaged over the center sites within $|{\bf r}_i|\leq 10 a$. As shown in Fig.~\ref{fig3}f, the value of $\delta n^2$ rapidly increases and then exhibits an irregular oscillation in the case without the phase imprinting, while it is almost constant {in the case that} the negative-temperature state is properly created by simultaneously achieving the maximum kinetic, interaction, and potential energies. 
%
\begin{figure}[t]
\includegraphics[scale=0.55]{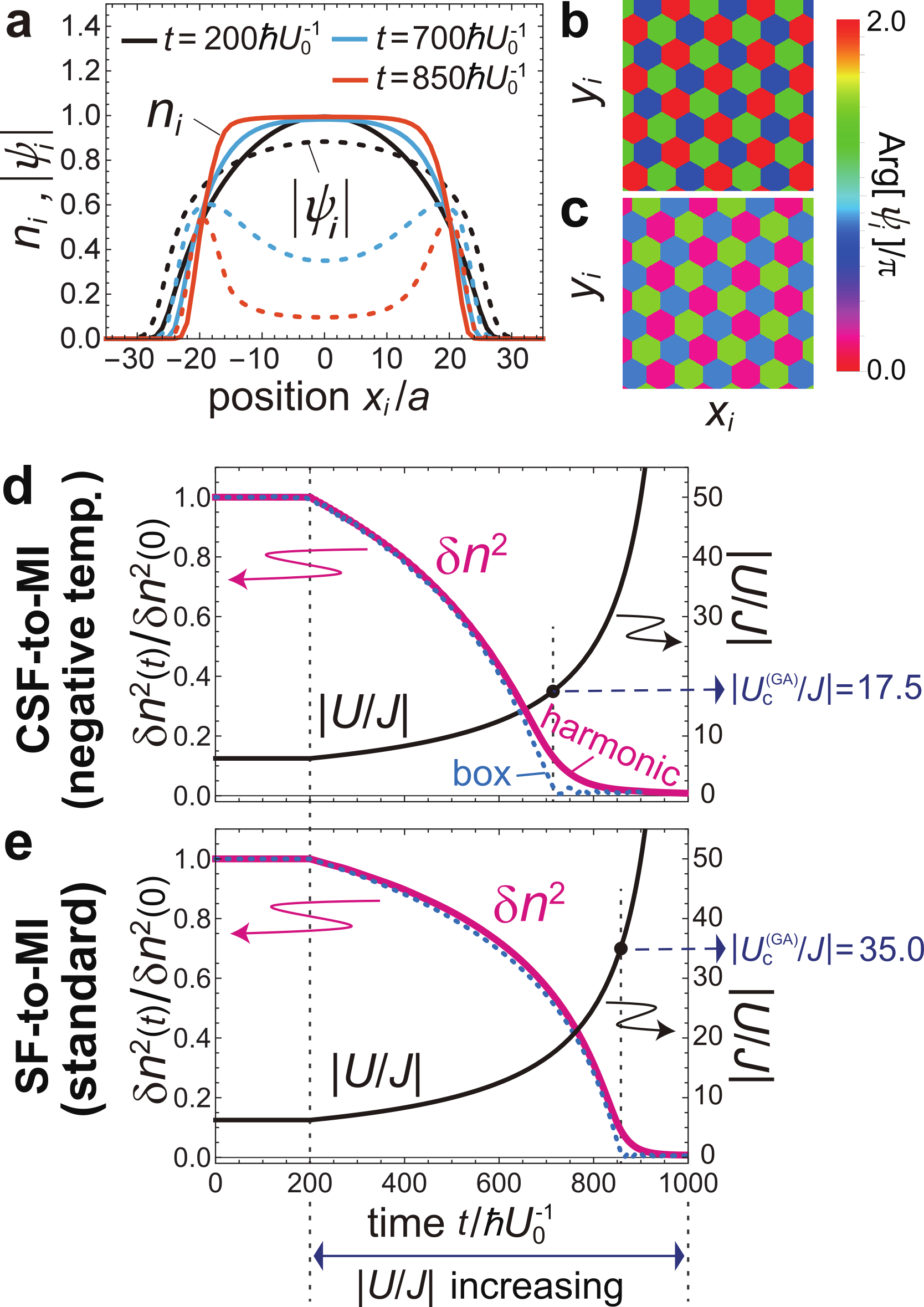}
\caption{\label{fig4}
{\bf Quantum phase transition from {chiral superfluid} (CSF) to {Mott insulator (MI)} within the {time-dependent Gutzwiller approach}.} (a) Time evolution of the profiles of density {$n_i$} (solid lines) and order parameter {$|\psi_i|$} (dashed lines) {in the harmonic trap along the cut of $y_i=a/\sqrt{3}$ (with $a$ being the lattice constant) for} the negative-temperature CSF state when $|U/J|$ increases for $t>200U_0^{-1}$. (b,c) Color plots for the local phase near the center of the trap at {(b)} {$t = 100U_0^{-1}$ and {(c)} $700U_0^{-1}$}. {(d,e)} {Time evolution of the density fluctuation $\delta n^2(t)/\delta n^2(0)$ for {(d)} a negative-temperature CSF state and {(e)} an unfrustrated {superfluid (SF)} state. The red solid and blue dashed lines represent the cases of harmonic and box-shaped trap potentials, respectively.} The time schedule for increasing {the ratio of the interaction and the hopping, $|U/J|$,} is plotted together. {The time $t$ is measured in units of the inverse of the initial interaction strength $U_0^{-1}$}}
\end{figure}
\\ \\ {\bf The CSF-to-MI phase transition.}
For the stable negative-temperature CSF state shown in Fig.~\ref{fig3}c, we slowly increase the value of $|U/J|$ for $t > 200 U_0^{-1} $ to observe the CSF-to-MI transition. In experiments, the tuning of $|U/J|$ is performed by controlling the height of the optical lattice~\cite{greiner-02}. Here, we assume that $|U/J|$ increases simply through $J$ decreasing linearly with $J=0.08U_0-0.0001(t U_0-200)U_0$ for $t > 200 U_0^{-1} $. As shown in Fig.~\ref{fig4}a, {when $|U/J|$ increases,} the MI plateau is gradually formed and $|\psi_i|$ in the trap center decreases {towards zero.} Until the transition point, the three-color phase profile in the CSF state is properly kept within a global phase shift ({Figs.~\ref{fig4}b and~\ref{fig4}c}).

To show the transition process more clearly, we plot in {Fig.~\ref{fig4}d} the time evolution of the scaled value of the density fluctuation $\delta{n}^2(t)/\delta{n}^2(0)$. {There we see that $\delta{n}^2(t)/\delta{n}^2(0)$} decreases {with} $|U/J|$ and almost vanishes at $|U/J|\approx |U^{({\rm GA})}_{\rm c}/J|=17.5$, which is consistent with the GA prediction of the critical point for the frustrated system. In the case of a box-shaped trap potential~\cite{gaunt-13,chomaz-15}, which is modeled by $V=0$ and the open boundary at $|{\bf r}_i|=36a$, the transition is more sharply observed (the blue dashed line in {Fig.~\ref{fig4}d} for $N=4692$). We also plot in {Fig.~\ref{fig4}e} the unfrustrated case of the standard SF-MI transition as a reference, which shows a sharp difference from the frustrated case in the values of $|U/J|$ of the transition region.

{At the end of this section, let us briefly mention the validity and limitation of the TDGA method with respect to the calculations presented above. The mean-field analysis with the site-decoupling approximation reproduces the exact wave function in the weak-coupling ($U\approx 0$) and strong-coupling ($U\rightarrow \infty$) limits, and is thus expected to reasonably interpolate the two limits in two or higher dimensions~\cite{GA1,zwerger-03,lewenstein-07}. In fact, the TDGA simulation has been often used to describe the dynamics of Bose-Hubbard systems~\cite{altman-05,snoek-07,snoek-12,saito-12}, which should be fairly reliable, at least, in the absence of strong quantum correlations, e.g., in the deep-SF regime or in three dimensions (3D)~\cite{mun-07}. It is therefore expected that the above TDGA simulation gave a proper description for the dynamical stability of the negative-temperature CSF state after the quench in the deep-SF regime. On the other hand, the site-decoupling treatment could fail to capture some quantitative features in the vicinity of the CSF-MI transition, including the location of the critical point and the values of critical exponents, since the intersite quantum correlations become important.}
\\ \\ {\bf Quantitative analysis by CMF+S with DMRG.}
We provide more quantitative estimation of the CSF-MI critical point for $V=0$ beyond the site-decoupling approximations (GA and TDGA) to predict the interplay effect of frustration and quantum fluctuations as a guideline necessary for experiments. The quantum effects on frustrated Bose gases in 2D lattices have been poorly studied due to the lack of effective computational methods. Here, we generalize and apply the numerical CMF+S method with 2D DMRG solver, which recently established in studies on frustrated quantum spins~\cite{yamamoto-19}, to the present system. 
We consider the $N_{\rm C}$-site cluster Hamiltonian on a triangular-shaped cluster (see the illustrations in Fig.~\ref{fig5}a) under the mean-field boundary condition. We work in the {twisted} frame, $\tilde{b}_{i}\equiv e^{-i{\bf Q}\cdot{\bf r}_i} \hat{b}_{i}$, with optimizing ${\bf Q}=(Q_x,0)$ (see Methods). The CMF+S method permits the systematic inclusion of quantum intersite correlations by increasing $N_{\rm C}$, which connects between the GA ($N_{\rm C}=1$) and {exactly quantum} ($N_{\rm C}\rightarrow \infty$) results.

\begin{figure}[t]
\includegraphics[scale=0.42]{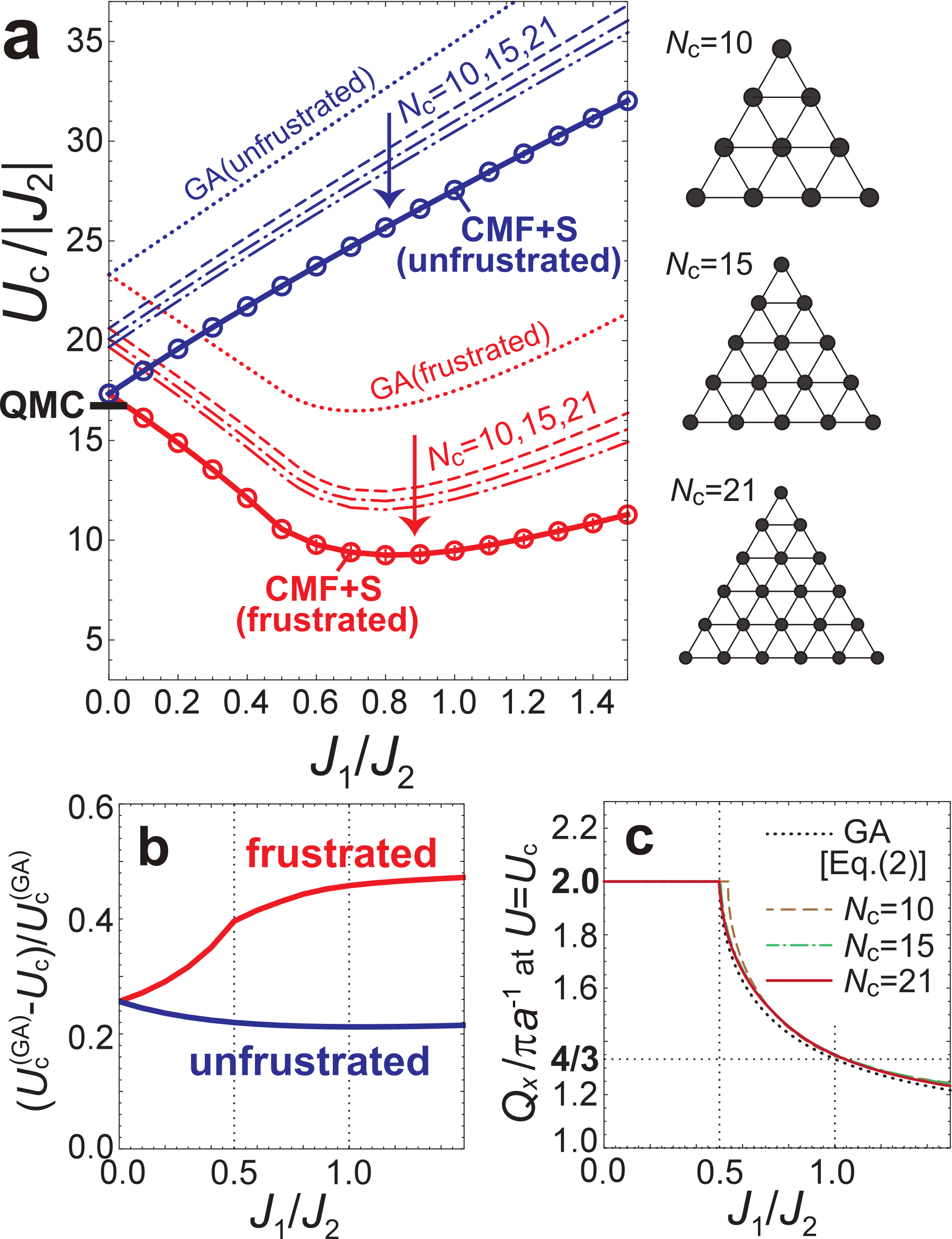}
\caption{\label{fig5}
{\bf Quantitative analysis on the quantum critical point.} (a) The {critical points $U_{\rm c}/|J_2|$ between the chiral superfluid (CSF) or superfluid (SF) state and the Mott insulator (MI) as functions of the hopping anisotropy $J_1/J_2$. The values obtained by the cluster mean-field plus scaling (CMF+S) analysis are compared with those obtained by the Gutzwiller approach (GA). In the CMF+S results, the error bars estimated from the variation in the linear fittings for different pairs of the $N_{\rm C}=10,15,21$ data are smaller than the symbol size.} The bar on the vertical axis marks the quantum Monte-Carlo {result} {on the square lattice~\cite{sansone-08} as reference}. (b) The relative difference between the GA and CMF+S results for the critical point. (c) The momentum {of the Bose-Einstein condensates, ${\bf Q}=(Q_x,0)$, in units of $\pi a^{-1}$ (with $a$ being the lattice constant) at the critical interaction strength} $U=U_{\rm c}$.  }
\end{figure}
Figure~\ref{fig5}a summarizes the $N_{\rm C}=10,15,21$ data and the CMF+S ($N_{\rm C}\rightarrow \infty$) result for the $\rho=1$ CSF-MI (SF-MI) critical point in the frustrated (unfrustrated) case with $J_1,J_2<0$ ($J_1,J_2>0$). 
The value of $U_{\rm c}/|J_2|$ for the frustrated case exhibits a nonmonotonic behavior with a dip around $J_1/J_2\approx 0.8$, while the unfrustrated one simply increases as the total hopping $J_1+2J_2$ increases. This is indeed the frustration effect, which destabilizes the CSF state. Besides, the value of $U_{\rm c}$ is strongly reduced from the GA prediction $U_{\rm c}^{({\rm GA})}$ due to the inclusion of the intersite quantum correlations. It should be noted that the relative difference $(U_{\rm c}^{({\rm GA})}-U_{\rm c})/U_{\rm c}^{({\rm GA})}$ between the GA and CMF+S values is much larger for the frustrated case ($\sim 40$-$50\%$) than the unfrustrated case ($\sim 20\%$) as shown in Fig.~\ref{fig5}b. This indicates that the quantum effects are strongly enhanced by the interplay with frustration. Figure~\ref{fig5}c shows that the BEC momentum $Q_x$ in the CSF state is little affected by the inclusion of quantum correlations. The slight variance of $Q_x$ from $4\pi/3$ at $J_2/J_1=1$ is thought to be due to the finite cluster-size effect (see Methods).
\\ \\ {\bf Detection method.}
The BEC momentum ${\bf Q}$ in the CSF state and the transition to the MI can be simply detected by time-of-flight (TOF) images of momentum distributions~\cite{greiner-02}. A more precise determination of the critical point can be made by {extracting the condensate fraction from the TOF images~\cite{spielman-08}, by observing the critical velocity using a moving optical lattice~\cite{mun-07}}, or by measuring the density fluctuation $\delta{n}^2$ using the quantum-gas microscope~\cite{mazurenko-17,parsons-16,boll-16,cheuk-16,hilker-17,brown-17}. The frustrated CSF-MI transition is easily {distinguishable} from the standard SF-MI transition thanks to the sharp difference in the value of $U_{\rm c}/|J_2|$. 
\\ \\    
{\bf \large Discussion}\\
We made a proposal and provided the necessary theoretical analysis for analog quantum simulation of frustrated quantum magnetism by using ultracold Bose gases in triangular optical lattices. We proposed an experimental protocol to create a frustrated quantum state at negative absolute temperature by performing a phase imprinting together with sudden inversion of the interatomic interaction and the trap potential. 
Simulating the time evolution, we demonstrated that a dynamically-stable superfluid state with chiral symmetry breaking was indeed realized, and underwent the quantum phase transition to the MI as the hopping amplitude decreased. Moreover, we performed state-of-the-art numerical calculations on the quantum critical point as a function of the spatial hopping anisotropy, which predicted a significant interplay of frustration and quantum fluctuations.
{The quantum dynamics in the presence of such enhanced fluctuations near the criticality is out of reach with the currently available numerical methods, and thus gives a strong motivation for future experimental quantum simulations as an interesting and important subject to be investigated.}

A connection of the present synthetic system to real materials of frustrated antiferromagnets can be obtained by using the approximate mapping from the bosonic to spin-1 operators~\cite{altman-02}: $\hat{b}_i\rightarrow \sqrt{\frac{\rho}{2}}\hat{S}_i^-,~\hat{n}_i\rightarrow \hat{S}_i^z+\rho$, which is valid in the vicinity of the transition between the CSF and MI phases at integer fillings $\rho$. The Hamiltonian~(\ref{hamiltonian}) is mapped onto the spin-1 {\rm XY} model
\begin{eqnarray}
\hat{\mathcal{H}}_{\rm spin}=-\sum_{i,j}J^{\rm XY}_{ij}(\hat{S}_{i}^x\hat{S}_{j}^x+\hat{S}_{i}^y\hat{S}_{j}^y) +D\sum_{i}(\hat{S}_i^z)^2 \label{hamiltonianS}
\end{eqnarray}
with the XY spin exchange $J^{\rm XY}_{ij}=-\rho J_{ij}$ and the single-ion anisotropy $D=U/2$. Therefore, the physics discussed in the present study is deeply related to the pressure-induced phase transition from the ``$120^\circ$'' magnetic order (corresponding to CSF) to nonmagnetic state (MI) in spin-1 easy-plane triangular antiferromagnets such as CsFeCl$_3$~\cite{kurita-16,hayashida-18}, which has recently attracted considerable attention in the connection with the novel excitations near quantum criticality~\cite{hayashida-19}. From the standpoint of fundamental statistical physics, the CSF-to-MI transition should be associated with the spontaneous U(1)$\times \mathbb{Z}_2$ symmetry breaking with respect to the global phase and the chirality determined by ${\bf q} ={\bf Q}$ or $-{\bf Q}$, whose quantum critical phenomena and universality class have not yet been established. The present bosonic system of synthetic antiferromagnets is advantageous for exploring exotic quantum critical phenomena in low dimensions, while the real materials have strong 3D couplings between the triangular layers~\cite{kurita-16,hayashida-18}.

Moreover, it has been expected that adding long-range interatomic interaction to the present system may give rise to an exotic chiral MI state~\cite{zaletel-14} in between possible separate U(1) and $\mathbb{Z}_2$ symmetry breakings. This is essentially equivalent to the so-called ``chiral liquid'' expected in a spin-1 frustrated magnet~\cite{wang-17}. Besides, our protocol for direct creation of a frustrated quantum state is advantageous for preparing the phases that are not neighboring to the MI phase, such as quantum spin liquids expected for $\rho = {\rm half}~{\rm integers}$~\cite{balents-10,eckardt-10}. Thus the present study gives a crucial guidepost for cold-atom quantum simulations of those exotic quantum frustrated physics. 
\\ \\    
{\bf \large Methods}\\
{\bf Triangular optical lattice with anisotropic hoppings.} 
A triangular optical lattice can be created by superposing three laser beams that intersect in the $x$-$y$ plane with wave vectors ${\bf  k}_1$ = $k_{\rm L} (1, 0)$, ${\bf  k}_2$ = $k_{\rm L} (-1/2, -\sqrt{3}/2)$, and ${\bf  k}_3$ = $k_{\rm L} (-1/2, \sqrt{3}/2)$ and equal frequency $\omega_{\rm L}$~\cite{becker-10}. All beams are linearly polarized orthogonal to the plane and each has field strength $E_i$ ($i=1,2,3$). The total electric field is given by 
\begin{equation}
{\bf  E}_{\rm tot}({\bf  r}, t)=\sum_{i=1}^3E_i \cos ({\bf  k}_i\cdot {\bf  r}-\omega_{\rm L} t+\phi_i){\bf  e}_z. 
\end{equation}
The dipole potential generated by the electric field is proportional to its squared amplitude,
\begin{eqnarray}
V({\bf  r})&\propto& \left|{\bf  E}_{\rm tot}({\bf  r}, t)\right|^2\nonumber\\&=&\frac{E^2_1+E^2_2+E^2_3}{2}+E_2E_3\cos ({\bf  b}_2\cdot {\bf  r}-\phi_{23})\nonumber\\
&&+E_1E_3\cos (({\bf  b}_1-{\bf b}_2)\cdot {\bf  r}+\phi_{13})\nonumber\\
&&+E_1E_2\cos ({\bf  b}_1\cdot {\bf  r}+\phi_{12}) + A(t),
\end{eqnarray}
where ${\bf  b}_1={\bf  k}_1-{\bf  k}_2$, ${\bf  b}_2={\bf  k}_3-{\bf  k}_2$, $\phi_{ij}=\phi_i-\phi_j$, and $A(t)$ represents the terms dependent on time $t$. Since the frequency of light is quite large, only the time-averaged value of $|{\bf  E}_{\rm tot}|^2$ can affect atoms. All the terms in $A(t)$ oscillate at frequency $2\omega_{\rm L}$, and thus can be dropped. Finally, we obtain a periodic dipole potential
\begin{eqnarray}
V({\bf  r})&=&-V_1 \cos ({\bf  b}_2\cdot {\bf  r}-\phi_{23})\nonumber\\
&&-V_2 \cos (({\bf  b}_1-{\bf b}_2)\cdot {\bf  r}+\phi_{13})\nonumber\\&&-V_3 \cos ({\bf  b}_1\cdot {\bf  r}+\phi_{12})+~{\rm const.},
\label{Vr}
\end{eqnarray}
where $V_1$, $V_2$, and $V_3$ are proportional to $E_2E_3$, $E_1E_3$, and $E_1E_2$, respectively. A variation of the phases $\phi_i$ of the laser beams yield only a global shift of the lattice in position. The primitive lattice vectors ${\bf a}_1$ and ${\bf a}_2$ are given so that ${\bf a}_i\cdot{\bf b}_j=2\pi\delta_{ij}$. We define ${\bf a}_3=-{\bf a}_1-{\bf a}_2$ for convenience. For red-detuned lasers with $V_i>0$, the potential minima form a geometrically equilateral triangular lattice in lattice constant: $a=|{\bf a}_i|=4\pi/3k_{\rm L}=2\lambda_{\rm L}/3$ with $\lambda_{\rm L}$ being the laser wavelength. The spatial anisotropy in the hopping amplitudes can be introduced by the difference in $V_1$, $V_2$, and $V_3$ through tuning the laser intensities $E_1$, $E_2$, and $E_3$. For example, the set of the field strengths with the relation of $E_1=1.6 E_2=1.6 E_3$ ($1.6 V_1= V_2=V_3$) yields the potential landscape shown in Fig.~\ref{fig1}a, which gives anisotropy of ``isosceles-type'' in the nearest-neighbor hopping amplitudes. 
\\ \\
{\bf Tilting triangular optical lattice with an additional 1D optical lattice.}
To perform the phase-imprinting process in the protocol proposed here, one has to introduce a single-particle energy difference $\delta E$ between adjacent two sites. This could be directly implemented by a magnetic field gradient, which introduces an extra linear gradient potential. Here, let us also provide another way to perform the phase-imprinting process for preparing the commensurate ${\bf Q}={\bf Q}_{\rm M}\equiv (2\pi/a,0)$ (two-color) and ${\bf Q}={\bf Q}_{\rm K}\equiv (4\pi/3a,0)$ (three-color) states by the use of an additional 1D optical lattice. We suppose that a potential of 1D optical lattice is created with additional laser beams in the ${\bf a}_1$ direction:
\begin{eqnarray}
V_{\rm 1D}(x)=V^\prime \cos^2 (k_{\rm L}^\prime x +\phi^\prime)
\label{Vr1D}
\end{eqnarray}
with amplitude $V^\prime$, wave vector $k_{\rm L}^\prime=2\pi/\lambda_{\rm L}^\prime$, and phase $\phi^\prime$. Here, let us set $\phi_i=0$ ($i=1,2,3$) for the triangular optical lattice without loss of generality.

\begin{figure*}[t]
\includegraphics[scale=0.5]{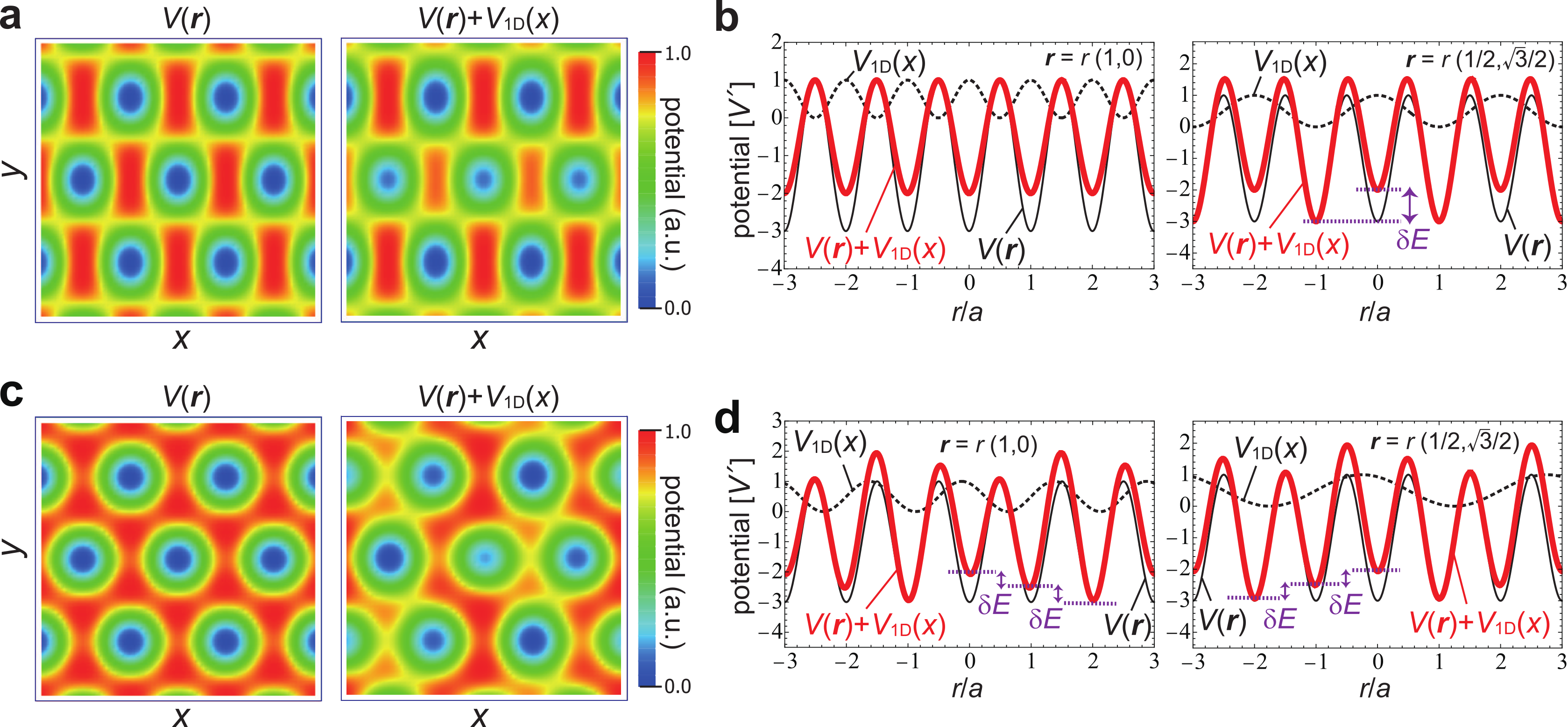}
\caption{\label{fig6}
{\bf Periodically-tilted triangular optical lattice.} (a) Potential landscapes of triangular optical lattice without and with an additional 1D periodic potential {(whose amplutude and phase are $V^\prime$ and $\phi^\prime$, respectively) for $2 V_1= V_2=V_3=V^\prime$ and $\phi^\prime=0$. The amplutudes of the three standing waves forming the triangular optical lattice are denoted by $V_1$, $V_2$, and $V_3$. The cuts along ${\bf r}=r(1,0)$ and ${\bf r}=r(1/2,\sqrt{3}/2)$ are shown in (b). The spatial coordinate ${\bf r}=(x,y)$ and its norm $r$ are measured in units of the lattice constant $a$. (c) and (d) Same as (a) and (b), respectively, for $V_1= V_2=V_3=V^\prime$ and $\phi^\prime=\pi/12$. }}
\end{figure*}
Let us first consider the range of $0\leq J_1/J_2\lesssim 0.5$ ($V_1\gg V_2=V_3$), in which the configuration of the local phase factor is expected to form the two-sublattice (say A and B) structure with the pitch vector ${\bf Q}={\bf Q}_{\rm M}$ as illustrated in the left panel of Fig.~\ref{fig2}a. For creating the local phase configuration by the phase imprinting, it is required to introduce a temporary single-particle energy difference $\delta E$ only between the two-sublattice groups of sites for time $\delta t=2\pi/\delta E$. This can be achieved by using an additional 1D optical lattice of magic wavelength defined by $\lambda_{\rm L}^\prime =4\lambda_{\rm L}/3$ and phase shift $\phi^\prime=0$ or $\pi/2$ (mod $\pi$). {Figures~\ref{fig6}a and~\ref{fig6}b show} an example of the total potential $V({\bf  r})+V_{\rm 1D}(x)$ with the parameters $2 V_1= V_2=V_3=V^\prime$ and $\phi^\prime=0$. Note that the two options in $\phi^\prime$ correspond to the exchange of A and B.

In a similar way, using a 1D periodic potential with $\lambda_{\rm L}^\prime =2\lambda_{\rm L}$, one can also implement a temporary energy difference $\delta E$ between the three-sublattice groups of sites, say A, B, and C. The additional lasers are shined for time $\delta t=4\pi/3\delta E$ to imprint the three-color phase configuration illustrated in the middle panel of Fig.~\ref{fig2}a to the initial SF state. {Figures~\ref{fig6}c and~\ref{fig6}d show} an example of the total potential $V({\bf  r})+V_{\rm 1D}(x)$ with the parameters $V_1= V_2=V_3=V^\prime$ and $\phi^\prime=\pi/12$. Note that the phase shift $\phi^\prime$ has six options, $(2n-1)\pi/12$ $(n=1,2,\cdots,6)$, reflecting the possible permutation of A, B, and C. 
\\ \\
{\bf The GA analysis for finite-momentum BEC states.}
Within the site-decoupling mean-field approximation, known as the GA, the effective local Hamiltonian at site $i$ is given by
\begin{eqnarray}
\hat{\mathcal{H}}_i^{\rm GA}&=&-\sum_{j\neq i}J_{ij}\left({\psi_j}\hat{b}_{i}^\dagger+\psi_j^\ast\hat{b}_{i}-\psi_{i}^\ast{\psi_j}\right)-\mu \hat{n}_{i}\nonumber\\&& +\frac{U}{2}\hat{n}_{i}(\hat{n}_{i}-1)+\frac{V}{a^2} |{\bf r}_i|^2 \hat{n}_{i},\label{MFhamiltonian}
\end{eqnarray}
as the result of the decoupling $\hat{b}_i^\dagger\hat{b}_j\approx \psi_j\hat{b}_i^\dagger+\psi_i^\ast \hat{b}_j-\psi_i^\ast\psi_j$ in the original Hamiltonian (1). 
The results displayed in Fig.~\ref{fig2}b are calculated under the assumption of finite-momentum BEC, $\psi_i= \bar{\psi} e^{i{\bf q}\cdot {\bf r}_i+\varphi}$, for $V=0$. In this case, one has only to consider a certain single site $i$, e.g., the site at the origin ${\bf r}_i={\bf 0}$, and $\varphi=0$ without loss of generality. Equation~(\ref{MFhamiltonian}) becomes 
\begin{eqnarray}
\hat{\mathcal{H}}_i^{\rm GA}=\varepsilon_{{\bf q}}\bar{\psi}\left(\hat{b}_{i}^\dagger+\hat{b}_{i}-\bar{\psi}\right)-\mu \hat{n}_{i} +\frac{U}{2}\hat{n}_{i}(\hat{n}_{i}-1),
\end{eqnarray}
for the origin site $i$. The minimization of the kinetic energy gives ${\bf q}={\bf 0}$ for $J_1,J_2>0$ and ${\bf q}={\bf Q}$ given in Eq.~(\ref{BECmomentum}) for $J_1,J_2<0$. The Hamiltonian $\hat{\mathcal{H}}_i^{\rm GA}$ can be easily diagonalized on the Fock state basis for the local wave function, $|\Psi_i\rangle\equiv \sum_{n=0}^{n_{\rm max}} f_i^{(n)}|n\rangle$, in which the maximum one-site occupation number $n_{\rm max}$ must be sufficiently large (we take $n_{\rm max}=10$). The order parameter is obtained from $\bar{\psi}=\sum_n \sqrt{n}f_i^{(n-1)\ast}f_i^{(n)}$ with eigenvector $f_i^{(n)}$ in a self-consistent way.

{Note that the effective one-body Hamiltonians $\hat{\mathcal{H}}_i^{\rm GA}$ in the unfrustrated and frustrated cases differ only in $\varepsilon_{{\bf q}}$ ($\varepsilon_{{\bf 0}}$ or $\varepsilon_{{\bf Q}}$). Therefore, if the values of all the other terms are multiplied by $|\varepsilon_{{\bf Q}}/\varepsilon_{{\bf 0}}|$, the results of the GA calculations, such as the value of $|\psi_i|$, for the unfrustrated system become exactly same as those for the frustrated system, except for the chiral phase distribution $e^{i{\bf Q}\cdot {\bf r}_i}$ and the overall energy scale (which is also multiplied by $|\varepsilon_{{\bf Q}}/\varepsilon_{{\bf 0}}|$).}

In the presence of the trap potential $V\neq 0$, the mean field $\psi_i$ can no longer be assumed to have spatially uniform amplitude. Therefore, one has to deal with Eq.~(\ref{MFhamiltonian}) on the entire lattice sites, each of which is connected to the six neighboring sites through the mean fields $\{\psi_{i\pm{\bf a}_n}; n=1,2,3\}$. To prepare the initial state shown in Fig. \ref{fig3}a, we solve the set of self-consistent equations $\psi_i=\sum_n \sqrt{n}f_i^{(n-1)\ast}f_i^{(n)}$ for all sites within a cutoff length $l_{\rm c}$ from the trap center. We take $l_{\rm c}=36a$. {The chemical potential $\mu$ is determined from the global number equation $\sum_i \langle \hat{n}_i\rangle =N$, which must be solved simultaneously with the equations for $\psi_i$. To efficiently achieve the convergence in the self-consistent calculations, we take the uniform solution for $\psi_i$ obtained in the absence of the trap potential as the initial condition, and employ the Newton-Raphson method. }
\\ \\
{\bf The TDGA simulation in a trap potential.}
The TDGA equation is given by
\begin{eqnarray}
i\frac{\partial}{\partial t}|\Psi_i(t)\rangle=\hat{\mathcal{H}}_i^{\rm GA}|\Psi_i(t)\rangle.\label{TDGA1}
\end{eqnarray}
From Eq.~(\ref{MFhamiltonian}) with $|\Psi_i(t)\rangle= \sum_{n} f_i^{(n)}(t)|n\rangle$, Eq.~(\ref{TDGA1}) becomes
\begin{eqnarray}
i\dot{f}_i^{(n)}(t)&=&-\sum_{j\neq i}J_{ij}\left(\sqrt{n}{\psi_j(t)}f_i^{(n-1)}(t)\right.\nonumber\\
&&\left.+\sqrt{n+1}\psi_j^\ast(t)f_i^{(n+1)}(t)-\psi_{i}^\ast(t){\psi_j}(t){f}_i^{(n)}(t)\right)\nonumber\\&&+\left(\frac{U}{2}(n-1)+\frac{V}{a^2} |{\bf r}_i|^2-\mu\right)n{f}_i^{(n)}(t)\label{TDGA2}
\end{eqnarray}
with $\psi_j(t)= \langle \Psi_j(t)|\hat{b}_j|\Psi_j(t)\rangle=\sum_n\sqrt{n}{f}_j^{(n-1)\ast}{f}_j^{(n)}(t)$.

We numerically solve the set of equations of motion for $f_i^{(n)}(t)$ using the Crank-Nicolson method~\cite{javanainen-99}. We first rewrite Eq.~(\ref{TDGA2}) in a matrix form:
\begin{eqnarray}
i\dot{{\bf f}}_i(t)&=&{\bf H}^{\rm GA}_i(\{\psi_j(t)\}){\bf f}_i(t),\label{TDGA3}
\end{eqnarray}
where ${\bf f}_i$ is the vector with the components $f_i^{(n)}$ ($0\leq n\leq n_{\rm max}$) and ${\bf H}^{\rm GA}_i$ is the $(n_{\rm max}+1)\times (n_{\rm max}+1)$ matrix form of the GA local Hamiltonian, which is time-dependent via the order parameters $\psi_j(t)$ of neighboring sites of site $i$. The explicit components of ${\bf H}^{\rm GA}_i$ are easily obtained from Eq.~(\ref{TDGA2}). The value of ${{\bf f}}_i$ after time evolution in a short time $\Delta t$ is given by
\begin{eqnarray}
{\bf f}_i(t+\Delta t)&=&\frac{{\bf I}-i\Delta t{\bf H}^{\rm GA}_i(\{\psi_j(t)\})/2}{{\bf I}+i\Delta t{\bf H}^{\rm GA}_i(\{\psi_j(t)\})/2}{\bf f}_i(t),\label{CN}
\end{eqnarray}
where ${\bf I}$ is the identity matrix. Here $\Delta t$ must be sufficiently shorter than the time scale of the physics considered. The time evolution of the system is numerically calculated, step by step, with short time interval $\Delta t$. In order to calculate ${\bf f}_i(t+\Delta t)$ at site $i$, the values of $\psi_j(t)$ on its neighboring sites are required. Therefore, one has to calculate $\psi_i(t)=\sum_n\sqrt{n}{f}_j^{(n-1)\ast}{f}_j^{(n)}(t)$ and update ${\bf f}_i(t)$ according to Eq.~(\ref{CN}) in parallel for all sites (within $|{\bf r}_i|\leq l_{\rm c}$). The local particle number at time $t$ can be calculated by $n_i(t)=\sum_n n |f_i^{(n)}(t)|^2$.
\\ \\
\begin{figure}[t]
\includegraphics[scale=0.52]{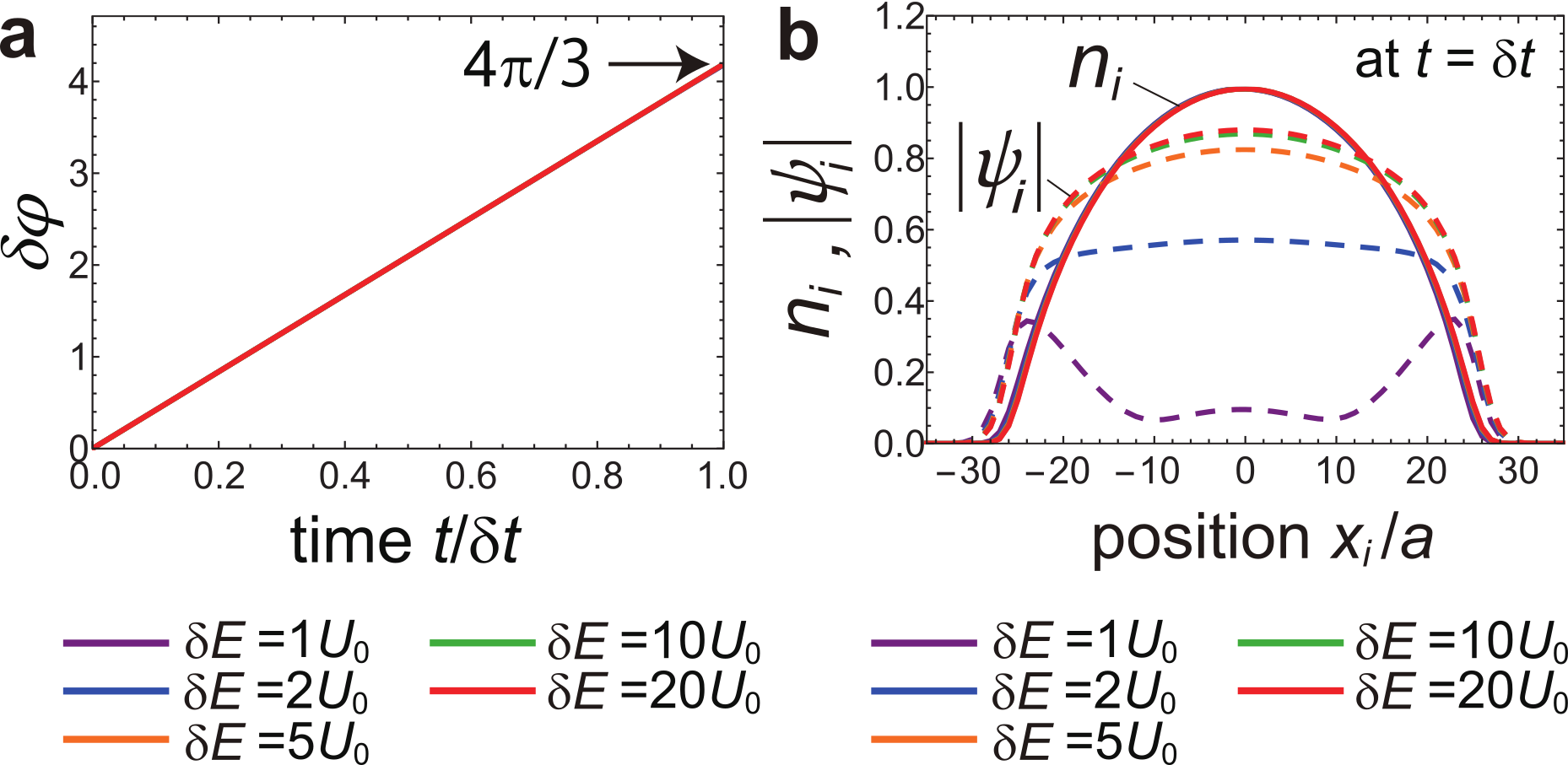}
\caption{\label{fig7}
{{\bf Phase imprinting.} (a) Time evolution of the phase difference between the neighboring sites in the ${\bf a}_1$ direction, $\delta\varphi\equiv \overline{({\rm Arg}[\psi_i]-{\rm Arg}[\psi_{i+{\bf a}_1}])}$, averaged over the center sites within $|{\bf r}_i|\leq 10 a$, in the presence of a temporary linear gradient potential. The colors correspond to different potential strengths $\delta E/U_0=1,2,5,10$, and $20$, although all the curves are almost overlapped. (b) The density (solid lines) and order-parameter (dashed lines) profiles in the final state after applying a gradient potential for time $\delta t$. }}
\end{figure}
{\bf The phase imprinting operation.} 
In the section ``TDGA simulation,'' we assume that the phase imprinting can be theoretically implemented by the operation $\sum_n f_i^{(n)}|n\rangle \rightarrow \sum_n e^{in{\bf Q}_{\rm K}\cdot{\bf r}_i } f_i^{(n)}|n\rangle$ on the local wave functions. Here, we provide a brief discussion on the experimental conditions for achieving such perfect phase imprinting. 

Let us consider the same initial state as shown in Fig.~\ref{fig3}a (with the parameters $N=1400$, $J_1=J_2=J=0.08U_0$, $U=U_0$, and $V=0.001U_0$) and its time evolution within the TDGA method in the presence of a temporary linear gradient potential $\hat{V}^{\rm ext}= \delta E \sum_i(x_i/a) \hat{n}_i$. As seen in Fig.~\ref{fig7}a, the chiral structure in the local phase, ${\rm Arg}[\psi_i]={\bf Q}\cdot {\bf r}_i$, could be successfully imprinted by applying the temporary potential for $\delta t=Q_x a/\delta E$ ($=4\pi/3\delta E$ for $J_1=J_2$). Figure~\ref{fig7}b indicates that the phase imprinting becomes almost perfect with no changes other than the local phase distribution when $\delta E$ exceeds $\sim 10U_0$. The corresponding imprinting time $\delta t\lesssim 0.4U_0^{-1}$ is much shorter than the typical time scale of the experiments on the SF(CSF)-MI transition (see Fig.~\ref{fig4}).
\\ \\
{\bf The CMF+S analysis with 2D DMRG solver for bosons.}
In the CMF+S analysis, we consider the $N_{\rm C}$-site cluster Hamiltonian
\begin{eqnarray}
\tilde{\mathcal{H}}_{\rm C}&=&-\sum_{i,j\in {\rm C}}J_{ij}e^{i{\bf q}\cdot({\bf r}_j-{\bf r}_i)}\tilde{b}_{i}^\dagger\tilde{b}_{j}+\frac{U}{2}\sum_{i\in {\rm C}}\hat{n}_{i}(\hat{n}_{i}-1)\nonumber\\
&&-\bar{\psi}\sum_{i\in \partial {\rm C}}\left(\sum_{j\notin {\rm C}}J_{ij}e^{i{\bf q}\cdot({\bf r}_j-{\bf r}_i)}\tilde{b}_{i}^\dagger+{\rm H.c.}\right)\label{hamiltonianC}
\end{eqnarray}
on a triangular-shaped cluster of $N_{\rm C}=10,15,21$ sites. The mean-field boundary condition on the cluster-edge sites $\partial {\rm C}$ is implemented by the third term, {and the twisted frame, $\tilde{b}_{i}\equiv e^{-i{\bf q}\cdot{\bf r}_i} \hat{b}_{i}$, is adopted}. 
The cluster Hamiltonian (\ref{hamiltonianC}) is treated with 2D DMRG solver. Here we take the maximum one-site occupation $n_{\rm max}=4$, which is confirmed to be sufficient for the discussion near the $\rho=1$ SF-MI (CSF-MI) transition. For large-size clusters and especially for the large Hilbert space of bosons, the exact diagonalization is practically not realistic as a solver for the cluster problem. Therefore, we employ the DMRG on the equivalent 1D chain model with long-range hoppings and mean fields (see Fig.~\ref{fig8}a). The DMRG calculation is performed in the standard way but with the mean-field terms in Eq.~(\ref{hamiltonianC})~\cite{yamamoto-19}. The dimension of the truncated matrix product states kept in the present DMRG calculations is typically $\sim 10^3$ to obtain numerically precise results. 
In order to solve 
\begin{eqnarray}
\bar{\psi}=\frac{1}{N_{\rm C}}\sum_{i\in {\rm C}}\langle \tilde{b}_{i} \rangle_{\tilde{\mathcal{H}}_{\rm C}(\bar{\psi})}
\end{eqnarray}
in a self-consistent way, we iteratively perform the DMRG calculations until convergence.

Note that when we put a real number $\bar{\psi}$ as an input for $\tilde{\mathcal{H}}_{\rm C}(\bar{\psi})$ in the fixed global gauge, the output $N_{\rm C}^{-1}\sum_{i\in {\rm C}}\langle \tilde{b}_{i} \rangle_{\tilde{\mathcal{H}}_{\rm C}(\bar{\psi})}$ includes a small but finite imaginary component ($\lesssim 4\%$ for $N_{\rm C}=21$). This is due to a finite-size effect; the order with uniform amplitude $\bar{\psi}$ and spiral phase twist $\exp [{i{\bf q}\cdot {\bf r}_i] }$ is not fully commensurate with the shape of the finite-size clusters with the mean-field boundary. We just ignore the small imaginary component in the calculations for each $N_{\rm C}$ since it decreases with $N_{\rm C}$ and is expected to vanish at the limit of $N_{\rm C}\rightarrow \infty$.

The optimization of the spiral twist ${\bf q}$ is performed in the following way: For different values of ${\bf q}=(q_x,0)$, the ``provisional'' critical point $U^\ast_{\rm c}/|J_2|$ (at which $\bar{\psi}=0^+$) is numerically determined (Fig.~\ref{fig8}b). The maximum value of $U^\ast_{\rm c}(q_x)/|J_2|$ and {the corresponding $(q_x,0)$} were adopted as the CMF+S prediction of the critical point $U_{\rm c}/|J_2|$ and the BEC momentum ${\bf Q}$ at the critical point, respectively. The slight variance of $Q_x$ from $4\pi/3$ at $J_1/J_2=1$ (see Fig.~\ref{fig5}c) is thought to stem from the same finite cluster-size effect mentioned above. 

\begin{figure*}[t]
\includegraphics[scale=0.4]{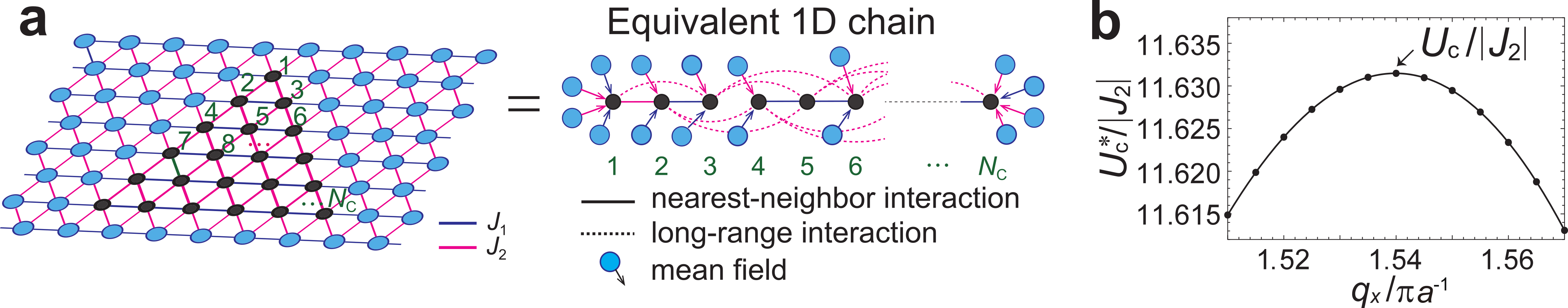}
\caption{\label{fig8}
{\bf Cluster mean-field calculations with 2D {density matrix renormalization group}.} (a) Mapping of the 2D cluster problem with the mean-field boundary onto an equivalent 1D chain with long-range interactions and mean fields. (b) Typical behavior of the provisional critical point $U^\ast_{\rm c}/|J_2|$ as a function of the given momentum ${\bf q}=(q_x,0)$ {in units of $\pi a^{-1}$} for {the hopping anisotropy} $J_1/J_2=0.7$ and {the cluster size} $N_{\rm C}=21$. }
\end{figure*}
\begin{figure}[t]
\includegraphics[scale=0.49]{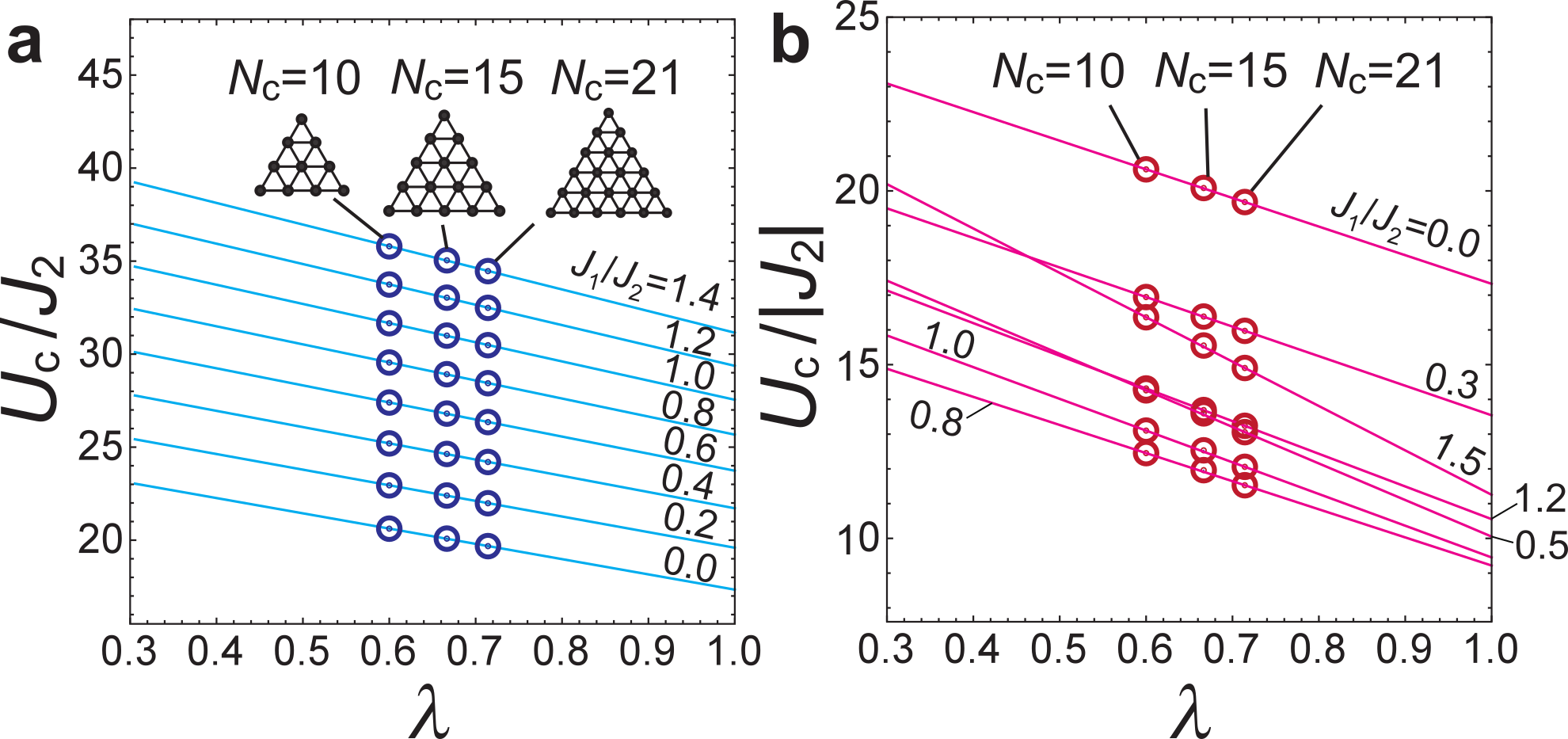}
\caption{\label{fig9}
{\bf Scaling analysis in the {cluster mean-field plus scaling analysis}.} Cluster-size scalings of the critical points for the transitions {(a) between the superfluid and Mott insulator states in the unfrustrated ($J_1,J_2>0$) case and (b) between the chiral superfluid and Mott insulator states in the frustrated ($J_1,J_2<0$) case}. The extrapolated ($\lambda\rightarrow 1$) values are plotted as a function of {the hopping anisotropy} $J_1/J_2$ in Fig.~\ref{fig5}a.}
\end{figure}
The CMF+S curves in Fig.~\ref{fig5}a are obtained from the size scaling of the phase boundaries for $N_{\rm C}=10,15,21$. Figure~\ref{fig9} shows the extrapolation of the $N_{\rm C}=10,15,21$ data to $N_{\rm C}\rightarrow \infty$ ($\lambda\rightarrow 1$) for several values of $J_1/J_2$ with a linear function of the scaling parameter $\lambda\equiv N_{\rm B}/3N_{\rm C}$~\cite{yamamoto-12-2,yamamoto-14}. Here, $N_{\rm B}$ is the number of NN bonds treated exactly in the cluster ($N_{\rm B}=18,30,45$ for $N_{\rm C}=10,15,21$, respectively). {The error bars in Fig.~\ref{fig5}a are estimated from the variation in the linear fittings for different pairs of the $N_{\rm C}=10,15,21$ data.} 
\\

{\bf \large Acknowledgements}\\
This work was supported by KAKENHI from Japan Society for the Promotion of Science: grant numbers 18K03525 (D.Y.), {19K03691 (D.Y.),} 19H01854 (T.F.), 18K03492 (I.D.), 18H05228 (I.D.), CREST from Japan Science and Technology Agency No.~JPMJCR1673 (D.Y. and I.D.), ImPACT Program of Council for Science, Technology and Innovation (Cabinet Office, Government of Japan) (T.F.), and Q-LEAP program of MEXT, Japan (I.D.). 

\end{document}